\documentclass[conference]{IEEEtran}

\usepackage[dvipsnames]{xcolor}
\usepackage{epsfig}
\usepackage{times}
\usepackage{float}
\usepackage{afterpage}
\usepackage{amsmath}
\usepackage{amstext,cite}
\usepackage{amssymb,bm}
\usepackage{latexsym}
\usepackage{color}
\usepackage{graphicx}
\usepackage{amsmath}
\usepackage{amsthm}
\usepackage{graphicx}
\usepackage[center]{caption}
\usepackage{pstricks}
\usepackage{caption}
\usepackage{subcaption}
\usepackage{booktabs}
\usepackage{multicol}
\usepackage{lipsum}
\usepackage[T1]{fontenc}
\usepackage{hyperref}
\usepackage{aecompl}
\usepackage{mathrsfs}

\usepackage{soul}
\usepackage{cancel}

\allowdisplaybreaks

\long\def\comment#1{}

\newcommand{\dv}{{\mathbf d}}

\newcommand{\Ac}{{\magenta \mathcal A}}

\newcommand{\Ec}{{\mathcal E}}

\newcommand{\Gc}{{\mathcal G}}

\newcommand{\Ic}{{\mathcal I}}
\newcommand{\Jc}{{\mathcal J}}

\newcommand{\Lc}{{\blue \mathcal L}}

\newcommand{\Qc}{{\mathcal Q}}

\newcommand{\Sc}{{\mathcal S}}
\newcommand{\Tc}{{\mathcal T}}

\newcommand{\Vc}{{\mathcal V}}

\newcommand{\qsf}{{\mathsf q}}

\newcommand{\Bsf}{{\mathsf B}}

\newcommand{\Ksf}{{\mathsf K}}

\newcommand{\Msf}{{\mathsf M}}
\newcommand{\Nsf}{{\mathsf N}}

\newcommand{\Rsf}{{\mathsf R}}

\providecommand{\definitionname}{Definition}

\title{A General Coded Caching Scheme \\ for Scalar Linear Function Retrieval}

\begin{document}

\author{
\IEEEauthorblockN{Yinbin Ma and Daniela Tuninetti,}
\IEEEauthorblockA{University of Illinois Chicago, Chicago, IL 60607, USA \\ Email:\{yma52, danielat\}@uic.edu}
}

\maketitle

\IEEEpeerreviewmaketitle

\begin{abstract}
Coded caching aims to minimize the network's peak-time communication load by leveraging the 
information pre-stored in the local caches at the users. 
The original single file retrieval setting by Maddah-Ali and Niesen has been recently extended to general Scalar Linear Function Retrieval (SLFR) by Wan {\it et al.},
who proposed a linear scheme that surprisingly achieves the same optimal load (under the constraint of uncoded cache placement) as in single file retrieval. 
This paper's goal is to characterize the conditions under which a general SLFR linear scheme is optimal and gain practical insights into why the specific choices made by Wan {\it et al.} work.
This paper shows that the optimal decoding coefficients are necessarily the product of two terms, one only involving the encoding coefficients and the other only the demands. In addition, the relationships among the encoding coefficients are shown to be captured by the cycles of certain graphs. Thus, a general linear scheme for SLFR can be found by solving a spanning tree problem.
\end{abstract}

\section{Introduction}
\label{sec:into}

Coded caching, originally introduced by Maddah-Ali and Niesen (MAN) in~\cite{maddah2014fundamental}, has been the focus of much research efforts recently as it predicts, for networks with a server delivering a single file to each cache-aided user, that it is possible to achieve a communication load that does not scale with the number of users.
Yu {\it et al.} in~\cite{yu2017exact} improved on the delivery phase of the MAN scheme by removing the MAN multicast message transmissions that are redundant when a file is requested by multiple users, and thus showed that the converse bound under the constraint of uncoded cache placement by Wan {\it et al.} in~\cite{wan2020index} is tight.
Wan {\it et al.} in~\cite{wan2020cache}
recently extended the MAN setup so as to allow users to request general scalar linear combinations of the files stored at the server. 
Despite the fact that the 
number of possible demands increases exponentially in the number of files,~\cite{wan2020cache} surprisingly showed that the optimal communication load is the same as for the single file retrieval setting, at least under uncoded cache placement. 

The scheme proposed in~\cite{wan2020cache} is linear. As in~\cite{yu2017exact}, the server selects of set of leaders (whose demand vectors are a linearly independent spanning set of the set of all possible demands) and creates multicast messages by performing linear combinations of demanded subfiles that were not cached; the coefficients for such linear combinations are referred to as {\it encoding coefficients} and can be optimized. As in~\cite{yu2017exact}, multicast messages that would only be useful for non-leader users are not sent and have to be locally reconstructed as linear combinations of sent multicast messages; the coefficients for such linear combinations are referred to as {\it decoding coefficients} and must guarantee that each user correctly decodes its demanded linear combination of files. The choice of encoding and decoding coefficients in~\cite{wan2020cache} is rather non trivial and not a simple extension of~\cite{yu2017exact}, which actually fails to guarantee successful decoding on finite fields of characteristics strictly larger than two. The encoding coefficients chosen in~\cite{wan2020cache}, inspired by private function retrieval in~\cite{sun2018capacity}, all have unit modulo but alternate in sign among leaders and among non-leaders. Such a choice works (with corresponding decoding coefficients given, up to a sign, by determinants of certain matrices derived from the demand matrix) but the reason why it is so could not be explained. 

This paper aims to gain insights into why the choices in~\cite{wan2020cache} work by analyzing the most general linear scheme (i.e., general encoding and decoding coefficients). Our main contribution is to show that the optimal decoding coefficients are necessarily the product of two terms, one only involving the encoding coefficients and the other only the determinants of certain matrices derived from the demands. In addition, we characterize the relationships the encoding coefficients need to satisfy in order to guaranteed successful decoding as cycles on certain graphs. {\it Thus, we show that a general SLFR linear scheme can be found by solving a spanning tree problem.}

The rest of the paper is organized as follow.
Section~\ref{sec:model} introduces the cache-aided scalar linear function retrieval (SLFR) problem and summarizes related work. 
Section~\ref{sec:main} presents our main result, which is proved in Section~\ref{sec:mainproof}.
Section~\ref{sec:conclusion} concludes the paper.
Some examples can be found in Appendix.

In this paper we use the following notation convention.
\begin{itemize}
  \item Calligraphic symbols denote sets, bold symbols 
  vectors, and sans-serif symbols 
  system parameters.
  \item $|\cdot|$ is the cardinality of a set or the length of a vector.
  \item $\mathrm{det}(M)$ is the determinant of the matrix $M$.
  \item $1_{\{\Ec\}}$ is the indicator function of the event $\Ec$.
  \item $M[\Qc, \Sc]$ is the submatrix of $M$ obtained by selecting the rows indexed by $\Qc$  and the columns indexed by $\Sc$.
  \item For an integer $b$, we let $[b] := \{1, \ldots, b\}$.
  \item For a ground set $\Gc$ and an integer $t$, we let $\Omega_{\Gc}^{t} := \{ \Tc \subseteq \Gc : |\Tc| = t\}$.
Moreover, $\Sc \setminus \Qc := \{k: k \in \Sc, k \notin \Qc\}$.
  \item $\mathsf{Ind}_{\Sc,k}$ returns the position of the element $k\in\Sc$, where the element of the integer set $\Sc$ are considered in increasing order. For example, $\mathsf{Ind}_{\{3,5\},3} = 1$ and $\mathsf{Ind}_{\{3,5\},5} = 2$. By convention $\mathsf{Ind}_{\Sc,k}=0$ if $k\not\in\Sc$.
\end{itemize}

\section{Problem Formulation and Known Results}
\label{sec:model}

\subsection{Problem Formulation}
A $(\Ksf,\Nsf,\qsf, \Msf, \Rsf)$ SLFR problem has one central server that has access to a library of $\Nsf$ files (denoted as $F_1,\ldots, F_{\Nsf}$), each of $\Bsf$ independent and uniformly distributed symbols over the finite field $\mathbb{F}_{\qsf}$, for some prime-power $\qsf$.
The server communicates through an error-free shared link at {\it load} $\Rsf$ to $\Ksf$ users, where each has a local {\it memory} to store up to $\Msf$ files.
The worst-case load $\Rsf^\star(\Msf), \Msf\in[0,\Nsf],$ for the SLFR problem is defined as in~\cite{wan2020cache}, which is not explicitly written here for same of space (as it also appears next). 

\subsection{Known Results}
It was shown in~\cite{wan2020cache} that requesting arbitrary scalar linear functions of the files from the server does not incur any load penalty compared to the case of requesting a single file, that is, the lower convex envelope of the following points is achievable 
\begin{align}
(\Msf, \Rsf)
= \left(\frac{\Nsf t}{\Ksf}, \frac{\binom{\Ksf}{t+1}-\binom{\Ksf- \min(\Nsf,\Ksf) }{t+1}}{\binom{\Ksf}{t}}  \right), \forall t\in [0:\Ksf]. 
\label{eq:YMA worst load}
\end{align}
Moreover, the tradeoff in~\eqref{eq:YMA worst load} is optimal among all schemes with uncoded cache placement~\cite{yu2017exact,wan2020cache} and to within a factor two otherwise~\cite{yu2018characterizing}. The scheme  in~\cite{wan2020cache} is as follows.

\paragraph{Cache Placement}
Partition the position indices as
\begin{align}
[\Bsf] = \left\{ \Ic_\Tc : \Ic_\Tc \subseteq [\Bsf], \Tc \in \Omega_{[\Ksf]}^{t}, \  |\Ic_\Tc| = \Bsf/\binom{\Ksf}{t} \right\}, 
\label{eq:position partition}
\end{align}
and define (with a Matlab-like notation) the sub-files as
\begin{align}
F_{i,\Tc} := F_i(\Ic_\Tc) \in \mathbb{F}_{\qsf}^{\Bsf/\binom{\Ksf}{t}}, \ \forall \Tc\in \Omega_{[\Ksf]}^{t}, \ \forall i\in[\Nsf].
\label{eq:new file split}
\end{align}

The cache of user $k\in[\Ksf]$ is populated as
\begin{align}
Z_k = \{ F_{i,\Tc} : \Tc \in \Omega_{[\Ksf]}^{t}, k\in\Tc, i\in[\Nsf] \} \in \mathbb{F}_{\qsf}^{\Bsf \Nsf \binom{\Ksf-1}{t-1}/\binom{\Ksf}{t}}.
\label{eq:cachecontent}
\end{align}
The memory size is thus 
$\Msf = \Nsf \binom{\Ksf-1}{t-1}/\binom{\Ksf}{t} = \Nsf \ t/\Ksf$
as in~\eqref{eq:YMA worst load}.

\paragraph{Delivery}
The demand of user $k \in [\Ksf]$ is represented by the row vector $\dv_k=(d_{k,1}, \ldots, d_{k,\Nsf}) \in \mathbb{F}_{\qsf}^{\Nsf}$, meaning that he 
needs to successfully retrieve the scalar linear function (i.e., operations are element-wise across files)
\begin{align}
B_{k}  
 := d_{k,1} F_1 +\ldots+ d_{k,\Nsf}  F_{\Nsf} \in \mathbb{F}_{\qsf}^{\Bsf}, \ \forall k\in[\Ksf].
\label{eq:LFtoR}
\end{align} 
As for the sub-files, define the {\it demand-blocks} as 
\begin{align}
B_{k,\Tc}  = B_{k}(\Ic_\Tc)
\in \mathbb{F}_{\qsf}^{\Bsf/\binom{\Ksf}{t}}, 
\ \forall \Tc \in \Omega_{[\Ksf]}^{t}, \ \forall k\in[\Ksf].
\label{eq:block}
\end{align}

Some demand-blocks can be computed based on the cache content available locally at the users in~\eqref{eq:cachecontent}, while the remaining ones need to be delivered by the server. 
Let $\mathbb{D}:=[\dv_1;\ldots;\dv_{\Ksf}] \in \mathbb{F}_{\qsf}^{\Ksf \times \Nsf}$ be the {\it demand matrix}.
Let $\Lc\subseteq[\Ksf]$ such that $\text{rank}_{\qsf}(\mathbb{D}) = \text{rank}_{\qsf}(\mathbb{D}[\Lc,:]) = |\Lc| =: r$ be the {\it leader set}, which is not unique but its size is (as every finite-dimensional vector space has a basis).
Let $\mathbb{D}^\prime\in \mathbb{F}_{\qsf}^{\Ksf \times |\Lc|}$ denote the {\it transformed demand matrix} defined as
\begin{align}
[\mathbb{D}^\prime]_{k,\ell} = \begin{cases}
1_{\{k = \ell\}} & \text{if} \ k\in\Lc \\
x_{k,\ell}       & \text{if} \ k\not\in\Lc \\
\end{cases}, 
\ \forall k\in[\Ksf], \ \forall \ell\in\Lc, 
\label{eq:transformed demand matrix}
\end{align}
i.e., the demand-blocks of non-leaders in~\eqref{eq:LFtoR} are expressed as a linear combination of the demand-blocks of the leaders as
\begin{align}
  B_{k,\Tc} &= \sum_{\ell\in\Lc} x_{k,\ell} \ B_{\ell,\Tc}, 
  \forall \Tc \in \Omega_{[\Ksf]}^{t}, \ \forall k\in[\Ksf]\backslash\Lc,
\label{eq:Bk for nonLeaders}
\end{align}
where the existence of the coefficients $\{x_{u,\ell} \in \mathbb{F}_\qsf : u\in \overline{\Lc}, \ell \in \Lc \}$ in~\eqref{eq:Bk for nonLeaders} follows from linear algebra.
The server forms the following multicast messages 
\begin{align}
W_{\Sc}
   &= \sum_{k\in \Sc} \alpha_{k,\Sc\setminus \{k\}} \ B_{k,\Sc\setminus \{k\}} \in \mathbb{F}_{\qsf}^{\Bsf/\binom{\Ksf}{t}}, 
   \ \forall \Sc \in \Omega_{[\Ksf]}^{t+1}, 
\label{eq:function retrieval delivery on Fq KAI's choice}
\end{align}
for some  {\it encoding coefficients} 
\begin{align}
\{ \alpha_{k,\Sc\setminus \{k\}} \in \mathbb{F}_{\qsf} \setminus\{0\}: \ k\in[\Ksf], \Sc \in \Omega_{[\Ksf]}^{t+1} \}.
\label{eq:enc-coeffs}
\end{align}

The server sends all multicast messages in~\eqref{eq:function retrieval delivery on Fq KAI's choice} that are useful for the leaders, that is, $X\in \mathbb{F}_{\qsf}^{\Delta+\Bsf(\binom{\Ksf}{t+1}-\binom{\Ksf-|\Lc|}{t+1})/\binom{\Ksf}{t}}$ for
\begin{align}
X = \{W_{\Sc} :  \Sc \in \Omega_{[\Ksf]}^{t+1}, |\Sc\cap\Lc|>0 \} \cup \{ \Lc, \mathbb{D}^\prime \}.
\label{eq:sent}
\end{align}
Note that sending the chosen leader set and the transformed demand matrix requires $\Delta = |\Lc|\lceil \log_\qsf(\Ksf) \rceil+\Ksf+|\Lc|$ symbols,
where $\Delta$ does not scale with the file length $\Bsf$.
The worst-case load is for $r = |\Lc|= \min(\Ksf,\Nsf)$ and equals
$\Rsf$
in~\eqref{eq:YMA worst load}.

For a given $\Sc \in \Omega_{[\Ksf]}^{t+1}$,  user $k\in\Sc$ can decode the missing demand-block $B_{k,\Sc\setminus \{k\}}$ from $W_{\Sc}$.
The multicast messages $\{W_{\Ac} : \Ac \in \Omega_{[\Ksf]\setminus \Lc}^{t+1} \}$ must be locally reconstructed from the transmitted ones 
in~\eqref{eq:sent} so that each user can recover all its missing demand-blocks.
For $\Ksf - r \geq t+1$, we seek to express 
\begin{align}
W_{\Ac} 
  = \sum_{\Sc \in \Omega_{[\Ksf]}^{t+1}, |\Sc\cap\Lc|>0 } \beta^{(\Ac)}_{\Sc} \ W_{\Sc}, \ \forall\Ac \in \Omega_{[\Ksf]\setminus \Lc}^{t+1},
\label{eq:whatweseek}
\end{align}
by an appropriate choice of the {\it decoding coefficients} 
\begin{align}
\{ \beta^{(\Ac)}_{\Sc} \in \mathbb{F}_{\qsf}: \Sc \in \Omega_{[\Ksf]}^{t+1}, |\Sc\cap\Lc|>0, \ \Ac \in \Omega_{[\Ksf]\setminus \Lc}^{t+1}  \}.
\label{eq:dec-coeffs}
\end{align} 
The choice of decoding coefficients must work for all realizations of the demand-blocks\footnote{
The leader set $\Lc$, the encoding coefficients in~\eqref{eq:enc-coeffs} and the decoding coefficients in~\eqref{eq:dec-coeffs}  are a function of $\mathbb{D}$ in general; such a dependency is not made explicit here in order not to clutter the notation. 
}.

In~\cite{wan2020cache} it was proposed that in~\eqref{eq:function retrieval delivery on Fq KAI's choice} one alternates between $\pm 1$ the encoding coefficients as
\begin{align}
\alpha_{k,\Sc\setminus\{k\}} =  (-1)^{\mathsf{Ind}_{\Sc\cap \Lc,k}+\mathsf{Ind}_{\Sc\setminus \Lc,k}},
\forall k\in\Sc.
\label{eq:Kai general separation}
\end{align}
which results in decoding coefficients that are equal, up to a sign, to determinants of certain sub-matrices of $\mathbb{D}^\prime$ in~\eqref{eq:transformed demand matrix}.
A reason for the choice of alternating signs in~\eqref{eq:Kai general separation} (and the resulting decoding coefficients) was not given in~\cite{wan2020cache}. The open question is whether such a choice is fundamental.

We answer this open question by analyzing a general linear scheme in the form of~\eqref{eq:function retrieval delivery on Fq KAI's choice} and~\eqref{eq:whatweseek}. We show that:
(1) the signs of the encoding coefficients must follow a pattern where they alternate, but not necessarily as in~\eqref{eq:Kai general separation}, and their modulo need not be one; 
(2) the decoding coefficients are proportional to the determinants of certain matrices obtained from the transformed demand matrix, but the proportionality coefficient need not have modulo one; and, finally and importantly, 
(3) the encoding and decoding coefficients must satisfy certain relationships that are captured by the cycles of a graph.

\section{Main Result}
\label{sec:main}
Our main result is to show that the linear scheme in~\eqref{eq:function retrieval delivery on Fq KAI's choice} and~\eqref{eq:whatweseek}
is correct if and only if the following holds.

The local reconstruction of non-sent multicast messages in~\eqref{eq:whatweseek} simplifies to solving
\begin{align}
0 &= \sum_{\Sc \in \Omega_{\Ac\cup\Lc}^{t+1}} \beta^{(\Ac)}_{\Sc} W_{\Sc} : \ \beta^{(\Ac)}_{\Ac} = -1, \ \forall \Ac \in \Omega_{[\Ksf]\setminus \Lc}^{t+1},
\label{eq:whatweseekNEW}
\end{align}
where in~\eqref{eq:whatweseekNEW} the summation is over subsets of $\Ac\cup\Lc$ 
(in total $\binom{|\Lc|+t+1}{t+1}$ terms in~\eqref{eq:whatweseekNEW}) rather than over some subsets of $[\Ksf]$ 
(in total $\binom{\Ksf}{t+1}-\binom{\Ksf-|\Lc|}{t+1}$ terms in~\eqref{eq:whatweseek}).
Eq\eqref{eq:whatweseekNEW} is solved, for any realization of the files, by using decoding coefficients 
\begin{subequations}
\begin{align}
\beta^{(\Ac)}_{\Sc}   &= \widetilde{\beta}^{(\Ac)}_{\Sc} \cdot \mathrm{det}\left(\mathbb{D}^\prime[\Ac\setminus\Sc, \Sc\setminus\Ac]\right),  
\\& \forall \Sc \in \Omega_{\Ac\cup\Lc}^{t+1}, \ \forall \Ac \in \Omega_{[\Ksf]\setminus \Lc}^{t+1},
\label{eq:opt beta}
\end{align}
\end{subequations}
where the part of the decoding coefficients that does not depend on the demands (denoted as $\widetilde{\beta}^{(\Ac)}_{\{k\} \cup \Tc}$ next) and the encoding coefficients (denoted as $\alpha_{k,\Tc}$ next) must satisfy
\begin{subequations}
\begin{align}
\widetilde{\beta}^{(\Ac)}_{\{k\} \cup \Tc} &\cdot  \alpha_{k,\Tc} = (-1)^{\phi^{(\Ac)}_{k, \Tc}} \cdot \mathsf{c}^{(\Ac)}_{\Tc},
\label{eq:opt alpha}
\\
  \phi^{(\Ac)}_{k, \Tc} 
    &= \begin{cases}
    1 + \mathsf{Ind}_{(\{k\} \cup \Tc) \setminus \Ac,k}  
        & k \in \Lc \setminus \Tc \\
    \mathsf{Ind}_{\Ac \setminus \Tc,k}  
        & k \in \Ac \setminus \Tc \\
    \end{cases},
\label{eq:opt sign}
\\ 
&\forall \Tc \in \Omega_{\Ac \cup \Lc}^{t}, \ \forall k\in (\Ac \cup \Lc) \setminus \Tc,
\end{align}
\label{eq:opt ALL}
\end{subequations}
\hspace{-4pt}for some constants $\{  \mathsf{c}^{(\Ac)}_{\Tc} \in \mathbb{F}_\qsf : \Tc \in \Omega_{\Ac \cup \Lc}^{t}, \ \forall k\in (\Ac \cup \Lc) \setminus \Tc \}$. 
Finally, the relationships in~\eqref{eq:opt ALL} can be represented on an undirected graph that has the $\widetilde{\beta}^{(\Ac)}_{\Sc}$'s and the $\mathsf{c}^{(\Ac)}_{\Tc}$'s  as vertices and whose edges are labeled by the encoding coefficients according to the constraints in~\eqref{eq:opt alpha}. {\it A spanning tree on such a graph identifies all the encoding coefficients that are free to vary}, in other words, cycles on such a graph identify constraints that the encoding coefficients must satisfy.

{\bf Remark.} The reason why the signs of the encoding coefficients (and the resulting decoding coefficients) must alternate in~\cite{wan2020cache} is because of the condition in~\eqref{eq:opt sign}, which is satisfied by the choice in~\eqref{eq:Kai general separation}; however the alternating patten in~\eqref{eq:Kai general separation} in just one possible feasible linear scheme. The choice of coefficients in~\eqref{eq:Kai general separation} (and the resulting decoding coefficients) has the following advantages:
(a) the scheme does not involve divisions other than by elements of unit modulo, which in turns allows one to extend the scheme to monomial retrieval as well~\cite{wan2020cache}; and
(b) the scheme works irrespective of the characteristics of the finite field.
$\hfill\square$

\section{Proof of Main Result}
\label{sec:mainproof}
We shall start to prove the result in Section~\ref{sec:main} from the case $\Ksf-|\Lc|=t+1$ in Section~\ref{sec:one non leader set only} (i.e., only the multicast message indexed by $\Ac=\overline{\Lc}$ must be reconstructed in~\eqref{eq:whatweseek}), then in Section~\ref{sec:many leader sets} we shall argue that the case $\Ksf-|\Lc| > t+1$ can be solved by analyzing several systems with only $|\Lc|+t+1$ users each. Moreover, we provide a complete characterization of all feasible linear schemes via graph theoretic properties. The proof holds for all $r=|\Lc|\in[\min(\Ksf,\Nsf)]$ and $t\in[0:\Ksf]$.

\subsection{Case $\Ksf-|\Lc|=t+1$}
\label{sec:one non leader set only}
We consider here a system with $\Ksf$ users, $r=|\Lc|$ leaders, and memory size parameterized by $t$, where $(t,r)$ are fixed and satisfy $\Ksf=r+t+1$.
For a subset $\Tc$ of $[\Ksf]$, we let $\overline{\Tc} := [\Ksf]\setminus\Tc.$ 
In particular, 
$\overline{\Lc}$ is the set of non-leader users.

Define the transformed demand matrix as in~\eqref{eq:transformed demand matrix}. 
Only the multicast message indexed by $\Ac=\overline{\Lc}$ needs to be reconstructed, thus for notation convenience we drop $\Ac$ from $\beta^{(\Ac)}_{\Sc}$ in~\eqref{eq:whatweseek}. We re-write~\eqref{eq:whatweseek}  with $\beta_{\overline{\Lc}}=-1$ (but actually any non-zero value will do), as follow
\begin{subequations}
\begin{align} 
&\mathbb{F}_{\qsf}^{\Bsf/\binom{\Ksf}{t}} \ni
0  = \sum_{\Sc \in \Omega_{[\Ksf]}^{t+1}} \beta_{\Sc} W_{\Sc}
 \\&= \sum_{\Sc \in \Omega_{[\Ksf]}^{t+1}} \beta_{\Sc} \sum_{k \in \Sc} \alpha_{k, \Sc \setminus \{k\}} \sum_{\ell \in \Lc} [\mathbb{D}^\prime]_{k,\ell} B_{\ell,\Sc \setminus \{k\}}
 \\&= \sum_{\Tc \in \Omega_{[K]}^{t}} \sum_{\ell \in \Lc} \sum_{k \in \overline{\Tc}} \beta_{\{k\} \cup \Tc} \ \alpha_{k, \Tc} \ [\mathbb{D}^\prime]_{k,\ell} B_{\ell,\Tc}. 
\end{align}
\label{eq:condition init}
\end{subequations}
\hspace{-4pt}Since~\eqref{eq:condition init} must hold for all $\{ B_{\ell,\Tc} \in \mathbb{F}_{\qsf}^{\Bsf/\binom{\Ksf}{t}} : \ell \in \Lc, \  \Tc \in \Omega_{[K]}^{t}\}$,  we equivalently rewrite it, $\forall \ell \in \Lc, \ \forall \Tc \in \Omega_{[K]}^{t}$,  as
\begin{subequations}
\begin{align}
\mathbb{F}_{\qsf}\ni
0  &= \sum_{k \in \overline{\Tc}} \beta_{\{k\} \cup \Tc} \ \alpha_{k, \Tc} \ [\mathbb{D}^\prime]_{k,\ell}  
 \\&= \sum_{k \in \overline{\Tc} \cap \Lc} \beta_{\{k\} \cup \Tc} \ \alpha_{k, \Tc} \ 1_{\{k=\ell\}}
 \\&+ \sum_{k \in \overline{\Tc} \cap \overline{\Lc}} \beta_{\{k\} \cup \Tc} \ \alpha_{k, \Tc} \ x_{k,\ell}, 
\end{align}
\label{eq:condition for all i and W}
\end{subequations}
\hspace{-5pt}by the dentition of transformed demand matrix in~\eqref{eq:transformed demand matrix}.
We finally rewrite~\eqref{eq:condition for all i and W} by separating it into two cases 
\begin{align}
\sum_{k \in \overline{\Tc} \cap \overline{\Lc} } \beta_{\{k\} \cup \Tc} \ \alpha_{k, \Tc} \ x_{k,\ell }
&= \begin{cases}
0    & \ell \in \Lc \cap \Tc, \\
-\beta_{\{\ell \} \cup \Tc} \ \alpha_{\ell, \Tc} &  \ell \in \Lc \cap \overline{\Tc} \\
\end{cases},
\label{eq:condition for all i and W again}
\notag \\ &\qquad \qquad \qquad \forall \Tc \in \Omega_{[K]}^{t}.
\end{align}

Next, we say that a set $\Tc \subseteq [\Ksf]$ is in `hierarchy~$h$' if $|\Tc\cap \Lc|=h$ for some $h\in[0:\min(|\Tc|,|\Lc|)]$.
We also say that $\beta_{\Sc}$ is in hierarchy~$h$ if $\Sc$ is in hierarchy~$h$.
{\it We next seek to show that in general the decoding coefficients in hierarchy~$h+1$ can be expressed as a linear combination of those in hierarchy~$h$.} 
\paragraph*{Initialization / hierarchy $h=1$}
$\beta_{\overline{\Lc}}=-1$ is the only decoding coefficient in hierarchy~$0$. 
By picking $\Tc = \overline{\Lc} \setminus \{u\}$, $u\in\overline{\Lc}$, and $\ell \in \Lc$ in~\eqref{eq:condition for all i and W again} (and thus $\overline{\Tc} \cap \overline{\Lc}=\{u\}$), we express the decoding coefficients in hierarchy~$1$ as follows 
\begin{align}
\beta_{ \{\ell\} \cup \overline{\Lc} \setminus \{u\} } 
= \frac{\alpha_{u, \overline{\Lc} \setminus \{u\}}}{\alpha_{\ell, \overline{\Lc} \setminus \{u\}}} \ x_{u,\ell},
\ \forall u\in\overline{\Lc}, \ \forall \ell\in\Lc.
\label{eq:H h+1 from h:init}
\end{align}

\paragraph*{Hierarchy $h$}
For any $\Tc \in \Omega_{[K]}^{t}$, from~\eqref{eq:condition for all i and W again} with $\ell\in\Tc$, 
\begin{align}
\sum_{k \in\overline{\Tc} \cap \overline{\Lc} } \beta_{\{k\} \cup \Tc} \ \alpha_{k, \Tc} \ x_{k,\ell} = 0, \ \forall \ell \in \Lc \cap \Tc.
\label{eq:H h+1 from nothing}
\end{align}

In particular, for a $\Tc$ in hierarchy $h>0$, we indicate WLOG (recall that here $|\overline{\Lc}|=\Ksf-r=t+1=|\Tc|+1$ and thus 
$|\Tc \cap \Lc| = h$, 
$|\Tc \cap \overline{\Lc}| = t-h$,
$|\overline{\Tc} \cap \Lc| = r-h$, 
$|\overline{\Tc} \cap \overline{\Lc}| = h+1$)
 \begin{align} 
 \Tc \cap \Lc &= \{ \ell_1, \ldots, \ell_h \} : \ell_1 < \ldots < \ell_h, \ \text{(leaders)},
 \label{eq:index W L}
 \\
 \overline{\Tc} \cap \overline{\Lc} &= \{ j_1, \ldots, j_h, j_{h+1} \} : j_1 < \ldots < j_{h+1}, 
 \label{eq:index CW CL}
\end{align}
and collect the $h$ constraints in~\eqref{eq:H h+1 from nothing} in matrix form as indicated in~\eqref{eq:H h+1 from nothing in matrix form} and~\eqref{eq:H h+1 from nothing in matrix form take2}, at the top of the next page, for all $\Tc \in \Omega_{[K]}^{t}$.
\begin{figure*}
\small
\begin{align}
   \begin{bmatrix}
         \beta_{\{j_1\} \cup \Tc} \ \alpha_{j_1,    \Tc} &
         \ldots &
         \beta_{\{j_{h}\} \cup \Tc} \ \alpha_{j_{h},\Tc} &
         \beta_{\{j_{h+1}\} \cup \Tc} \ \alpha_{j_{h+1},\Tc} 
    \end{bmatrix}
    \underbrace{
    \begin{bmatrix}
      x_{j_1, \ell_1} & \cdots & x_{j_1, \ell_h} \\
      \vdots & \ddots & \vdots \\
      x_{j_{h}, \ell_1} & \cdots & x_{j_{h}, \ell_h} \\
      x_{j_{h+1}, \ell_1} & \cdots & x_{j_{h+1}, \ell_h} \\
    \end{bmatrix}
    }_{ = \mathbb{D}^\prime[\overline{\Tc} \cap \overline{\Lc},\Lc \cap \Tc] \in \mathbb{F}_{\qsf}^{h+1 \times h} }
     = 
    0 \in \mathbb{F}_{\qsf}^{1 \times h},
\label{eq:H h+1 from nothing in matrix form}
\\
\begin{bmatrix}
         \frac{\beta_{\{j_1\} \cup \Tc} \ \alpha_{j_1,\Tc}}{\beta_{\{j_{h+1}\} \cup \Tc} \ \alpha_{j_{h+1},\Tc}} &
         \ldots &
         \frac{\beta_{\{j_h\} \cup \Tc} \ \alpha_{j_h,\Tc}}{\beta_{\{j_{h+1}\} \cup \Tc} \ \alpha_{j_{h+1},\Tc}} 
    \end{bmatrix}
    \underbrace{
    \begin{bmatrix}
      x_{j_1, \ell_1} & \cdots & x_{j_1, \ell_h} \\
      \vdots & \ddots & \vdots \\
      x_{j_h, \ell_1} & \cdots & x_{j_h, \ell_h}
    \end{bmatrix}
    }_{ = \mathbb{D}^\prime[\overline{\Tc} \cap \overline{\Lc} \setminus \{ j_{h+1} \},\Lc \cap \Tc] \in \mathbb{F}_{\qsf}^{h \times h} }
     = -
    \underbrace{
    \begin{bmatrix}
         x_{j_{h+1}, \ell_1} &
         \ldots &
         x_{j_{h+1}, \ell_h} 
    \end{bmatrix}
        }_{ = \mathbb{D}^\prime[\{ j_{h+1} \},\Lc \cap \Tc] \in \mathbb{F}_{\qsf}^{1 \times h}},
\label{eq:H h+1 from nothing in matrix form take2}
\end{align}
\end{figure*}
By Cramer's rule, the solution of~\eqref{eq:H h+1 from nothing in matrix form take2} 
can be written as
\begin{subequations}
\begin{align}
   &
(-1)^{h+1-i} \, \frac{\mathrm{det}\left(\mathbb{D}^\prime[\overline{\Tc} \cap \overline{\Lc} \setminus \{ j_i     \},\Lc \cap \Tc]\right)}
                           {\mathrm{det}\left(\mathbb{D}^\prime[\overline{\Tc} \cap \overline{\Lc} \setminus \{ j_{h+1} \},\Lc \cap \Tc]\right)}
\\&=\frac{\beta_{\{j_i\} \cup \Tc} \ \alpha_{j_i,\Tc}}{\beta_{\{j_{h+1}\} \cup \Tc} \ \alpha_{j_{h+1},\Tc}}, 
\forall i \in [h], \forall j_i \in \overline{\Tc} \cap \overline{\Lc},
\end{align}
\label{eq:H h+1 from nothing in matrix form take3}
\end{subequations}
\hspace{-4pt}or equivalently~\eqref{eq:H h+1 from nothing in matrix form take3} can be written as (recall $j\in\overline{\Tc} \cap \overline{\Lc}$)
\begin{subequations}
\begin{align}
&(-1)^1\frac{\beta_{\{j_1\} \cup \Tc} \ \alpha_{j_1,\Tc}}{{\mathrm{det}\left(\mathbb{D}^\prime[\overline{\Tc} \cap \overline{\Lc} \setminus \{ j_1     \},\Lc \cap \Tc]\right)}}
=
\ldots
\\&=
(-1)^{h+1}\frac{\beta_{\{j_{h+1}\} \cup \Tc} \ \alpha_{j_{h+1},\Tc}}{{\mathrm{det}\left(\mathbb{D}^\prime[\overline{\Tc} \cap \overline{\Lc} \setminus \{ j_{h+1}     \},\Lc \cap \Tc]\right)}}.
\end{align}
\label{eq:H h+1 from nothing in matrix form take4}
\end{subequations}
\hspace{-3pt}Notice that all the decoding coefficients in~\eqref{eq:H h+1 from nothing in matrix form take4} are in hierarchy $h$ if the set $\Tc$ is hierarchy $h$.

\paragraph*{Hierarchy $h+1$}
We plug the decoding coefficients in hierarchy $h$ from~\eqref{eq:H h+1 from nothing in matrix form take4} into~\eqref{eq:condition for all i and W again} with $\ell\in \overline{\Tc}$ and, by definition of determinant (i.e., Laplace expansion along a column), we obtain that for all $\Tc \in \Omega_{[K]}^{t}$ 
\begin{subequations}
\begin{align}
  &-\beta_{\{\ell\} \cup \Tc} \ \alpha_{\ell, \Tc}
  = \sum_{k \in \overline{\Tc} \cap \overline{\Lc} } \beta_{\{k\} \cup \Tc} \ \alpha_{k, \Tc} \ x_{k,\ell}
\label{eq:H h+1 from h:again 1}
\\&= \frac{\beta_{\{j_{h+1}\} \cup \Tc} \ \alpha_{j_{h+1},\Tc} }
                           {\mathrm{det}\left(\mathbb{D}^\prime[\overline{\Tc} \cap \overline{\Lc} \setminus \{ j_{h+1} \},\Lc \cap \Tc]\right)}
\\& \cdot \sum_{ i\in[h+1] } (-1)^{h+1-i} \, \mathrm{det}\left(\mathbb{D}^\prime[\overline{\Tc} \cap \overline{\Lc} \setminus \{ j_i     \},\Lc \cap \Tc]\right) \ x_{j_i,\ell}
\notag
\\&= (-1)^{h+1} \frac{\beta_{\{j_{h+1}\} \cup \Tc} \ \alpha_{j_{h+1},\Tc} } 
     {\mathrm{det}\left(\mathbb{D}^\prime[\overline{\Tc} \cap \overline{\Lc} \setminus \{ j_{h+1} \},\Lc \cap \Tc]\right)}
\\& \cdot (-1)^{-\mathsf{Ind}_{\Lc \cap \Tc \cup\{\ell\},\ell}}
  \mathrm{det}\left(\mathbb{D}^\prime[\overline{\Tc} \cap \overline{\Lc} ,\Lc \cap \Tc \cup\{\ell\}]\right),
\end{align}
\label{eq:H h+1 from h:again 3}
\end{subequations}
\hspace{-4pt}or equivalently, $\forall \Tc \in \Omega_{[K]}^{t}, \forall \ell \in \overline{\Tc} \cap \Lc,$  we have
\begin{align}
(-1)^{1+\mathsf{Ind}_{\Lc \cap \Tc \cup\{\ell\},\ell}} \frac{\beta_{\{\ell\} \cup \Tc} \ \alpha_{\ell, \Tc}}{\mathrm{det}\left(\mathbb{D}^\prime[\overline{\Tc} \cap \overline{\Lc} ,\Lc \cap \Tc \cup\{\ell\}]\right)}
= \text{eq\eqref{eq:H h+1 from nothing in matrix form take4}},
\label{eq:H h+1 from nothing in matrix form take5}
\end{align}
Notice that all the decoding coefficients in~\eqref{eq:H h+1 from nothing in matrix form take5} are in hierarchy $h+1$ if the set $\Tc$ is hierarchy $h$.

\paragraph*{Combing everything together}
We can interpret~\eqref{eq:H h+1 from nothing in matrix form take4} and~\eqref{eq:H h+1 from nothing in matrix form take5} as follows: for a set $\Tc \in \Omega_{[K]}^{t}$ and an element $k \in \overline{\Tc}$, we create a set $\Sc = \Tc \cup\{k\} \in \Omega_{[K]}^{t+1}$ that satisfies the following:
add a non-leader
\begin{subequations}
\begin{align}
k=j \in \overline{\Tc} \cap \overline{\Lc} 
 :& \ \overline{\Tc} \cap \overline{\Lc} \setminus \{ j  \} = \overline{\Lc} \setminus (\{j\} \cup \Tc),
\\& \ \Lc \cap \Tc = (\{j\} \cup \Tc) \setminus \overline{\Lc}, 
\end{align}
or add a leader
\begin{align}
k=\ell \in \overline{\Tc} \cap \Lc 
 :& \ \overline{\Tc} \cap \overline{\Lc} = \overline{\Lc} \setminus (\{\ell\} \cup \Tc),
\\& \ \Lc \cap \Tc \cup\{\ell\} = (\{\ell\} \cup \Tc) \setminus \overline{\Lc},
\end{align}
\end{subequations}
thus (recall $\overline{\Tc} \cap \overline{\Lc} = \overline{\Lc} \setminus \Tc$, $\overline{\Tc} \cap \Lc= \Lc \setminus \Tc$ and $\overline{\Tc} = [\Ksf] \setminus \Tc$)
\begin{subequations}
\begin{align}
&
\mathsf{c}^{(\overline{\Lc})}_{\Tc}
=
(-1)^{\phi^{(\overline{\Lc})}_{k, \Tc}} \alpha_{k, \Tc}
\cdot \widetilde{\beta}^{(\overline{\Lc})}_{\{k\} \cup \Tc},
\forall \Tc \in \Omega_{[K]}^{t}, \ \forall k\in \overline{\Tc}, 
\\& 
 \widetilde{\beta}^{(\overline{\Lc})}_{\{k\} \cup \Tc} :=
\frac{\beta_{\{k\} \cup \Tc} }{\mathrm{det}\left(\mathbb{D}^\prime[\overline{\Lc} \setminus (\{k\} \cup \Tc) , (\{k\} \cup \Tc) \setminus \overline{\Lc}]\right)},
\label{eq:H h+1 from nothing in matrix form take6 const}
\\&
\phi^{(\overline{\Lc})}_{k, \Tc} 
:= \begin{cases}
\mathsf{Ind}_{\overline{\Lc} \setminus \Tc,k}
   & k \in \overline{\Lc} \setminus \Tc, \\ 
{1 + \mathsf{Ind}_{(\{k\} \cup \Tc) \setminus \overline{\Lc},k}} 
   & k \in \Lc \setminus \Tc, \\ 
\end{cases},
\label{eq:H h+1 from nothing in matrix form take6 signs}
\end{align}
\label{eq:H h+1 from nothing in matrix form take6 ALL}
\end{subequations} 
\hspace{-4pt}for some constatns $\{\mathsf{c}^{(\overline{\Lc})}_{\Tc} : \Tc \in \Omega_{[K]}^{t} \}$.

The term 
in~\eqref{eq:H h+1 from nothing in matrix form take6 const} (that only depends on $\{k\} \cup \Tc$ as opposed to on both $k$ and $\Tc$) can be further expressed as a function of the encoding coefficients as follows.
For a set $\Sc \in \Omega_{[\Ksf]}^{t+1}, \Sc \neq \overline{\Lc},$ in hierarchy $h$ and by setting WLOG
\begin{align} 
  \Sc \cap \Lc &= \{ \ell_1, \ldots, \ell_{h} \} : \ell_1 < \ldots < \ell_{h}, \ \text{(leaders)}
  \label{eq:index S L}
  \\
  \overline{\Sc} \cap \overline{\Lc} &= \{ j_1, \ldots, j_{h} \} : j_1 < \ldots < j_{h}, \ \text{(non leaders)}
  \label{eq:index CS CL}
  \\
  \Sc \cap \overline{\Lc} &= \Jc,
  \
  \overline{\Lc} =  \{ j_1, \ldots, j_{h} \} \cup \Jc,
\end{align}
we iteratively use~\eqref{eq:H h+1 from h:again 3} to express $\beta_\Sc$ with $\Sc = \{\ell_1 \ldots \ell_{h}\} \cup \Jc$ as in~\eqref{eq:beta equationYM} at the top of the next page
\begin{figure*}[!t]
  \small
  \begin{subequations}
  \begin{align}
    \widetilde{\beta}^{(\overline{\Lc})}_{\{\ell_1 \ldots \ell_{h}\} \cup \Jc}
    &=\frac{\beta_{\{\ell_1 \ldots \ell_{h}\} \cup \Jc}}{\mathrm{det}\left( \mathbb{D}^\prime [ \overline{\Sc} \cap  \overline{\Lc}, \Sc \cap \Lc] \right)}
  = - \frac{\alpha_{j_{h}, \{\ell_1 \ldots \ell_{h-1}\} \cup \Jc}}{\alpha_{\ell_{h}, \{\ell_1 \ldots \ell_{h-1}\} \cup \Jc}} \ \frac{\beta_{\{j_h\} \cup \{\ell_1 \ldots \ell_{h-1}\} \cup \Jc}}{\mathrm{det}\left( \mathbb{D}^\prime [ \overline{\Sc} \cap  \overline{\Lc} \setminus \{j_h\}, \Sc \cap \Lc \setminus \{\ell_h\}] \right)}
  \\&= (-1)^h \ \frac{\alpha_{j_h, \{\ell_1 \ldots \ell_{h-1} \} \cup \Jc}}{\alpha_{\ell_h, \{\ell_{1} \ldots \ell_{h-1}\} \cup \Jc}} \ \frac{\alpha_{j_{h-1}, \{j_h\} \cup \{\ell_1 \ldots \ell_{h-2} \} \cup \Jc }}{\alpha_{\ell_{h-1}, \{j_h\} \cup \{\ell_1 \ldots \ell_{h-2} \} \cup \Jc }} \ldots \frac{\alpha_{j_{1}, \{j_h \ldots j_2\} \cup \Jc }}{\alpha_{\ell_{1},\{j_h \ldots j_2\} \cup \Jc}} \ \frac{\beta_{\{j_h \ldots j_1\} \cup \Jc}}{\mathrm{det}\left(\mathbb{D}^\prime [ \emptyset, \emptyset]\right)}
  \\&
    = (-1)^{h+1} \ \prod_{i=1}^h \frac{\alpha_{j_{i}, \{j_h \ldots j_{i+1}\} \cup \{\ell_1 \ldots \ell_{i-1} \} \cup \Jc }}{\alpha_{\ell_{i}, \{j_h \ldots j_{i+1}\} \cup \{\ell_1 \ldots \ell_{i-1} \} \cup \Jc }}, 
  \end{align}
  \label{eq:beta equationYM}
  \end{subequations}
  \end{figure*}
and where the last equality follows since by definition $\beta_{\{j_h \ldots j_1\} \cup \Jc} = \beta_{\overline{\Lc}} = -1$
and by convention $\mathrm{det}\left(\mathbb{D}^\prime [ \emptyset, \emptyset]\right)=1$.
Eq~\eqref{eq:beta equationYM} shows that each decoding coefficient is proportional to the determinant of a sub-matrix of the transformed demand matrix and that the proportionality coefficient (denoted as $\widetilde{\beta}^{(\overline{\Lc})}_{\{\ell_1 \ldots \ell_{h}\} \cup \Jc}$) depends only on the encoding coefficients; the encoding coefficients however are not all free to vary, as they need to satisfy the relationships imposed by~\eqref{eq:H h+1 from nothing in matrix form take6 const}.

\paragraph*{Graph representation}
The relationships among $\Vc_1 := \{\mathsf{c}^{(\overline{\Lc})}_{\Tc} : \Tc\in \Omega_{[K]}^{t}\}$ and $\Vc_2 := \{\widetilde{\beta}^{(\overline{\Lc})}_{\Sc} : \Sc\in \Omega_{[K]}^{t+1}\}$ imposed by~\eqref{eq:H h+1 from nothing in matrix form take6 ALL} can be represented by a graph.
We create an undirected graph $\Gc(\Vc, \Ec)$, where $\Vc:=\Vc_1 \cup \Vc_2$ is the vertex set and $\Ec := \{(\widetilde{\beta}^{(\overline{\Lc})}_{\{k\} \cup \Tc}, \mathsf{c}^{(\overline{\Lc})}_{\Tc}) : \Tc \in \Omega_{[K]}^{t}, k \in \overline{\Tc} \}$ is the edge set.
We assign label $(-1)^{\phi^{(\overline{\Lc})}_{k, \Tc}} \alpha_{k,\Tc}$ to edge $(\widetilde{\beta}^{(\overline{\Lc})}_{\{k\} \cup \Tc}, \mathsf{c}^{(\overline{\Lc})}_{\Tc}) \in \Ec$ to capture the relationship in~\eqref{eq:H h+1 from nothing in matrix form take6 ALL}.
We elect $\widetilde{\beta}_{\overline{\Lc}}$ to be the root node and assign to it the value $-1$ 
(but we could start from any other vertex with any non-zero value). 
We then create a spanning tree from that root\footnote{A spanning tree is a subset of the graph, which has {\it all the vertices of the graph covered with minimum possible number of edges}. Hence, a spanning tree does not have cycles and it cannot be disconnected. Moreover, every connected and undirected graph has at least one spanning tree.}. By doing so, we find values for all the vertices by using~\eqref{eq:H h+1 from nothing in matrix form take6 ALL}. 
One can easily see, by the properties of spanning trees, that the encoding coefficients on the edges of the spanning tree are free to vary (i.e., they can be  be any non-zero value), while the encoding coefficients  on edges that are not part of the spanning tree are determined through the following relationship: every path from the root to a node determines the value of the node by using~\eqref{eq:H h+1 from nothing in matrix form take6 ALL} and all those values must be equal;  in other words, every cycle in the graph, obtained by adding a edge that is not on the spanning tree to the spanning tree, is a constraint. 

This concludes the proof for the case $\Ksf-r = t+1$.

\begin{figure}
  \centering   
\vspace*{-2cm}
  \includegraphics[width=0.4\textwidth]{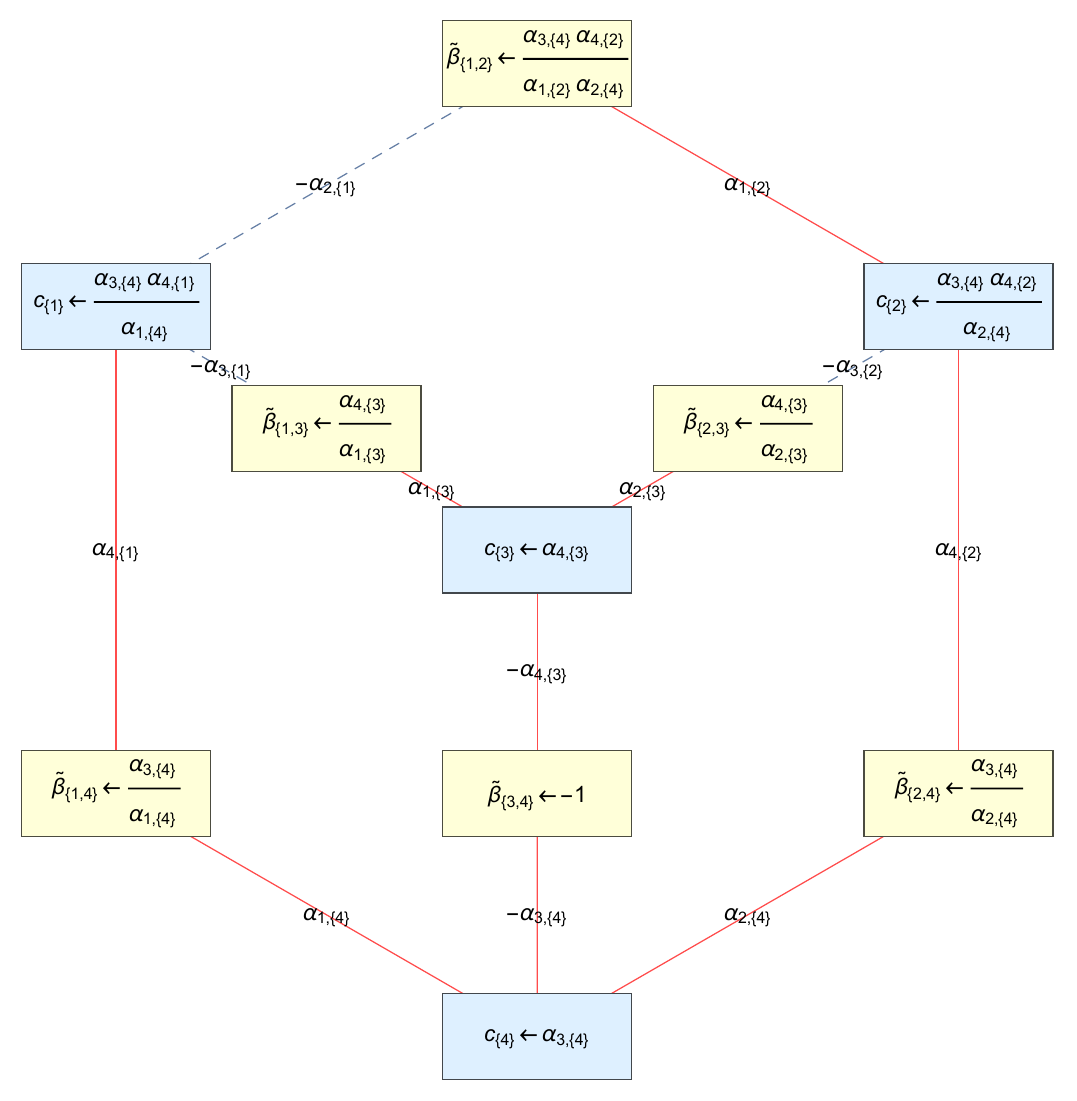} 
  \caption{\small 
  The graph and a possible spanning tree for the case $\Ksf=4, r=2, t=1$.
  For legibility, we removed the superscript $\overline{\Lc} = \{3,4\}$ from the vertices.
  The edges are labeled by an encoding coefficient with an appropriate sign.
  Solid edges form a spanning tree; the encoding coefficients on dotted edges are determined by using~\eqref{eq:H h+1 from nothing in matrix form take6 const}.
  The $\widetilde{\beta}$-vertexes are in a yellow box and the $\mathsf{c}$-vertexes in a cyan box;
  the expression on the RHS of the symbol $\leftarrow$ in a box is the value assigned to the vertex when we travel the graph from the root (i.e., $\widetilde{\beta}_{\{3,4\}}=-1$) along the spanning tree. 
  }
  \label{fig: k=4 r=2 t=1:spanning}
\end{figure}

\paragraph*{Example}
Fig.~\ref{fig: k=4 r=2 t=1:spanning} shows the described graph for the case of $K=4$ users, $r=2$ leaders, and memory size $t=1$ (i.e., each user can cache one file);
the edges of a possible spanning tree are marked by a solid red line;
the edges that are not in the spanning tree (doted blue line edges) correspond to the following constraints
\begin{subequations}
\begin{align}
\text{vertex} \ \mathsf{c}_1 : \alpha_{3,\{1\}} &= - \alpha_{1,\{3\}} \frac{\alpha_{4,\{1\}}}{\alpha_{1,\{4\}}} \frac{\alpha_{3,\{4\}}}{\alpha_{4,\{3\}}}, \label{eq:ex421:c1}\\
\text{vertex} \ \mathsf{c}_2 : \alpha_{3,\{2\}} &= - \alpha_{2,\{3\}} \frac{\alpha_{4,\{2\}}}{\alpha_{2,\{4\}}} \frac{\alpha_{3,\{4\}}}{\alpha_{4,\{3\}}}, \label{eq:ex421:c2}\\
\text{vertex} \ \widetilde{\beta}_{\{1,2\}} : \alpha_{2,\{1\}} &= - \frac{\alpha_{4,\{1\}}}{\alpha_{1,\{4\}}} \frac{\alpha_{4,\{2\}}}{\alpha_{2,\{4\}}} \alpha_{1,\{2\}}. \label{eq:ex421:c3}
\end{align}
\label{eq:ex421:wb12}
\end{subequations}
The relationships in~\eqref{eq:ex421:wb12} can arrived at by directly solving~\eqref{eq:whatweseek} as shown in Appendix~\ref{example: k=4 r=2 t=1}.

\subsection{Case $\Ksf-|\Lc| > t+1$}
\label{sec:many leader sets}
\begin{figure*}
  \centering
  \begin{subfigure}[b]{0.3\textwidth}
    \centering
    \includegraphics[width=\textwidth]{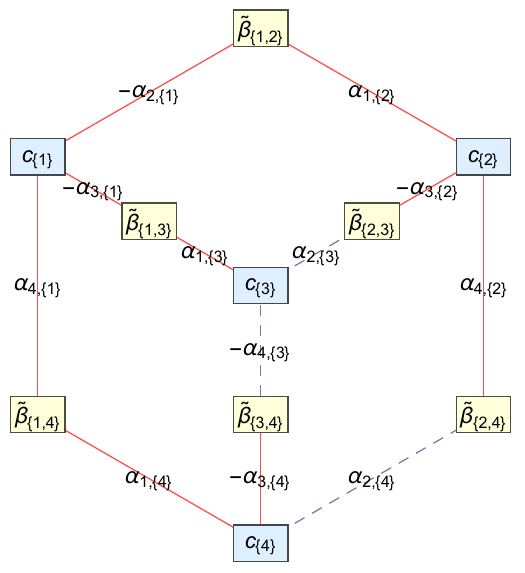}
    \caption{Component for $W_{\{3,4\}}$}
    \label{fig:521-w34}
  \end{subfigure}
  \begin{subfigure}[b]{0.3\textwidth}
    \centering
    \includegraphics[width=\textwidth]{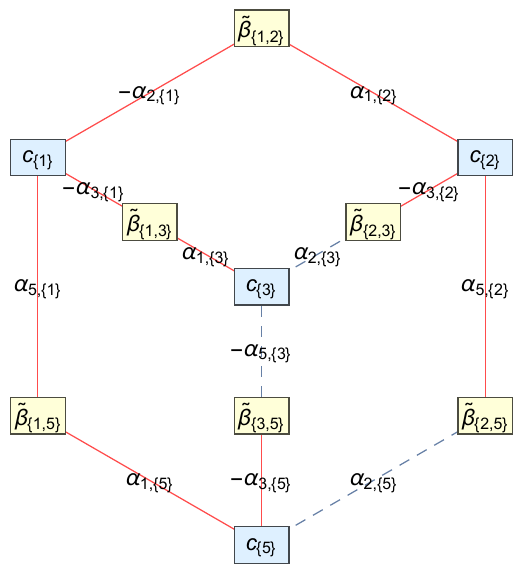}
    \caption{Component for $W_{\{3,5\}}$}
    \label{fig:521-w35}
  \end{subfigure}
  \begin{subfigure}[b]{0.3\textwidth}
    \centering
    \includegraphics[width=\textwidth]{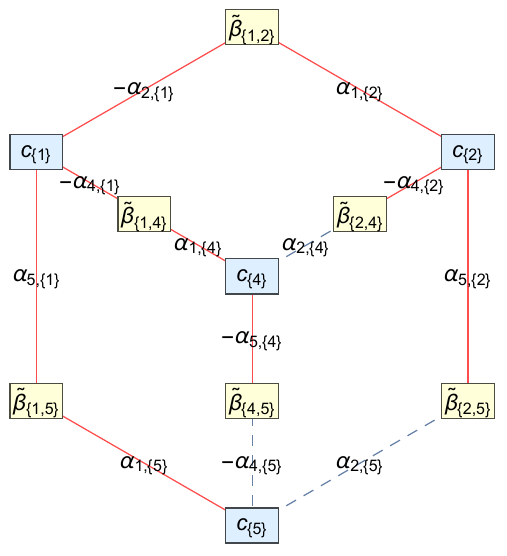}
    \caption{Component for $W_{\{4,5\}}$}
    \label{fig:521-w45}
  \end{subfigure}
  \caption{\small
  The graph and possible spanning trees for the case $\Ksf=5, r=2, t=1$.
  The convention is as in Fig.~\ref{fig: k=4 r=2 t=1:spanning}.
  For sake of legibility, we omitted the superscripts in the various sub-figures, which should be the index of the multicast message listed in the sub-caption.
  }
  \label{fig: k=5 r=2 t=1}
\end{figure*}

It is easy to see that in order to locally reconstruct all non-sent multicast messages as in~\eqref{eq:whatweseek} we need not sum over all sent multicast messages indexed by $\{\Sc \in \Omega_{[\Ksf]}^{t+1} : |\Sc\cap\Lc|>0\}$ but only on those indexed by $\{\Sc \in \Omega_{\Ac\cup\Lc}^{t+1}: \Sc \not= \Ac\}$. By doing so,  we can equivalently re-write~\eqref{eq:whatweseek} as in~\eqref{eq:whatweseekNEW}. In other words, for reconstructing multicast messages $W_{\Ac}$ we consider a ``reduced system'' with users in $\Ac\cup\Lc$
for which $W_{\Ac}$ is only multicast message to be reconstructed.
The analysis we did in Section~\ref{sec:one non leader set only}
applies to this ``reduced system'' with $|\Ac\cup\Lc| = t+1+r$ users.
After the substitutions $\Ac\cup\Lc$ instead of $[\Ksf]$, and $\Ac$ instead of $\overline{\Lc}$, the conditions in~\eqref{eq:H h+1 from nothing in matrix form take6 ALL} reads as stated in~\eqref{eq:opt ALL}.
\paragraph*{Graph representation}
The relationships in~\eqref{eq:opt ALL} can be represented on a graph as we did in Section~\ref{sec:main}. The resulting graph now has as many disconnects components as there are multicast messages to reconstruct. The edges of the various components are labeled by the encoding coefficients.
As an example, Fig.~\ref{fig: k=5 r=2 t=1} shows the graph and a set of possible spanning trees (one per disconnected component) for the case $\Ksf=5, r=2, t=1$, by using the same convention as in Fig.~\ref{fig: k=4 r=2 t=1:spanning}.
Unlike for the case $\Ksf-r=t+1$, here some encoding coefficients appear more than once in the graph, meaning that finding a spanning tree independently for each connect component may result in some encoding coefficients being part of one spanning tree (and thus being free to vary) while not being part of other spanning trees (and thus being determined by the corresponding `cycle' constraint). Since our goal here is to determined all encoding coefficients that are free to vary, we propose the following greed algorithm.

\begin{enumerate}
  \item We assign the ``priority score'' $1_{\{k \in \Lc\}} + 2|\Tc \cap \Lc|$ to encoding coefficient $\alpha_{k, \Tc}$, and sort all encoding coefficients in decreasing order of priority score. 
  \item We check each group of coefficients with the same priority score, and mark an encoding coefficient as ``free'' if the corresponding edges do not form a cycle with prior free coefficients in any of the components.
  \item We end after all coefficients have been checked. 
\end{enumerate}

The edges/encoding coefficients marked as ``free'' by this greedy algorithm are free to vary, as they are part of the spanning tree for each of the components in which they appear. 
The priority score aims to find the edges that are in the largest number of components at each step, and cycles are simultaneously broken in all components in order to build the spanning trees.
This greedy algorithm guarantees that the edges that are marked as ``not free'' (and are marked as such in every component they appear in) are in a cycle with the same set of ``free'' edges in all components they appear in, that is, although the same encoding coefficient appears to be constrained by multiple cycles, all those cycles involve edges with the same label and thus do not conflict. Appendix~\ref{example: k=5 r=2 t=1} explains the details of the greedy algorithm by directly solving~\eqref{eq:whatweseek}.

\section{Conclusion}
\label{sec:conclusion}
In this paper, we investigated the constraints that a linear scheme for cache-aided scalar linear function retrieval must satisfy in order to be feasible.
We showed that the constraints among the parameters of a feasible linear scheme are captured by the cycles of a certain graph.
Equivalently, we showed that 
a spanning tree for the graph identifies the parameters of the scheme that are free to vary. 
The structure of our general scheme sheds light into a scheme that had been previously proposed in the literature.
Ongoing work includes using similar ideas to explain the scheme in~\cite{sun2018capacity}.

This work was supported in part by NSF Award 1910309.

\appendices 

\section{Example: $\Ksf=4, r=2, t=1$}
\label{example: k=4 r=2 t=1}

WLOG, let $\Lc = \{1,2\}$ and thus $\overline{\Lc} = \{3,4\}$.
The multicast messages sent by the server are
\begin{align}
  W_{\{1,2\}} &= \alpha_{1,\{2\}}{\cyan B_{1,\{2\}}} + \alpha_{2,\{1\}}{\blue B_{2,\{1\}}}, \quad \text{(all leaders)},\\
  W_{\{1,3\}} &= \alpha_{1,\{3\}}B_{1,\{3\}} + \alpha_{3,\{1\}}B_{3,\{1\}}, \quad \text{(mixed)},\\
  W_{\{1,4\}} &= \alpha_{1,\{4\}}B_{1,\{4\}} + \alpha_{4,\{1\}}B_{4,\{1\}}, \quad \text{(mixed)},\\
  W_{\{2,3\}} &= \alpha_{2,\{3\}}B_{2,\{3\}} + \alpha_{3,\{2\}}B_{3,\{2\}}, \quad \text{(mixed)},\\
  W_{\{2,4\}} &= \alpha_{2,\{4\}}B_{2,\{4\}} + \alpha_{4,\{2\}}B_{4,\{2\}}, \quad \text{(mixed)}
\end{align}
and the multicast messages that is not sent is 
\begin{align}
W_{\{3,4\}} &= \alpha_{3,\{4\}}B_{3,\{4\}} + \alpha_{4,\{3\}}B_{4,\{3\}}, \quad \text{(all non leaders)}.
\end{align}

In order to reconstruct $W_{\{3,4\}}$ at users, we seek the decoding coefficients $\{\beta^{\{3,4\}}_{\Sc}: \Sc \in \Omega_{[4]}^2 , \Sc \neq \{3,4\}\}$ such that
\begin{align}
  W_{\{3,4\}}
        &= \beta^{\{3,4\}}_{\{1,2\}}W_{\{1,2\}} 
   \\&   + \beta^{\{3,4\}}_{\{1,3\}}W_{\{1,3\}}
         + \beta^{\{3,4\}}_{\{1,4\}}W_{\{1,4\}} 
   \\&   + \beta^{\{3,4\}}_{\{2,3\}}W_{\{2,3\}}
         + \beta^{\{3,4\}}_{\{2,4\}}W_{\{2,4\}},
\end{align}
that is, we aim to solve the following 
\begin{align}       
  & \alpha_{3,\{4\}}[x_{3,1}B_{1,\{4\}}+x_{3,2}B_{2,\{4\}}] + \alpha_{4,\{3\}}[x_{4,1}B_{1,\{3\}}+x_{4,2}B_{2,\{3\}}]
 \\    &= \beta^{\{3,4\}}_{\{1,2\}}\left( \alpha_{1,\{2\}}{\cyan B_{1,\{2\}}} + \alpha_{2,\{1\}}{\blue B_{2,\{1\}}} \right) 
 \\    &+ \beta^{\{3,4\}}_{\{1,3\}}\left( \alpha_{1,\{3\}}B_{1,\{3\}} + \alpha_{3,\{1\}}[x_{3,1}{\green B_{1,\{1\}}}+x_{3,2}{\blue B_{2,\{1\}}}] \right) 
 \\    &+ \beta^{\{3,4\}}_{\{1,4\}}\left( \alpha_{1,\{4\}}B_{1,\{4\}} + \alpha_{4,\{1\}}[x_{4,1}{\green B_{1,\{1\}}}+x_{4,2}{\blue B_{2,\{1\}}}] \right) 
 \\    &+ \beta^{\{3,4\}}_{\{2,3\}}\left( \alpha_{2,\{3\}}B_{2,\{3\}} + \alpha_{3,\{2\}}[x_{3,1}{\cyan B_{1,\{2\}}}+x_{3,2}{\magenta B_{2,\{2\}}}] \right) 
 \\    &+ \beta^{\{3,4\}}_{\{2,4\}}\left( \alpha_{2,\{4\}}B_{2,\{4\}} + \alpha_{4,\{2\}}[x_{4,1}{\cyan B_{1,\{2\}}}+x_{4,2}{\magenta B_{2,\{2\}}}] \right) 
 \end{align}
for any realization of the demand-blocks.
We thus equate the coefficients on the RRS and on the LRS of the above equation, as follows.

We start with the hierarchy~1 decoding coefficients
\begin{align}
  \text{for} \ B_{1,\{3\}} &: \quad \alpha_{4,\{3\}}x_{4,1}=\beta^{\{3,4\}}_{\{1,3\}}\alpha_{1,\{3\}} 
\\& \Longleftrightarrow  \frac{\beta^{\{3,4\}}_{\{1,3\}}}{x_{4,1}} = \frac{\alpha_{4,\{3\}}}{\alpha_{1,\{3\}}} = \widetilde{\beta}_{\{1,3\}}^{\{3,4\}}, \label{eq:apa b13}\\
  \text{for} \ B_{1,\{4\}} &: \quad \alpha_{3,\{4\}}x_{3,1}=\beta^{\{3,4\}}_{\{1,4\}}\alpha_{1,\{4\}}
\\& \Longleftrightarrow  \frac{\beta^{\{3,4\}}_{\{1,4\}}}{x_{3,1}} = \frac{\alpha_{3,\{4\}}}{\alpha_{1,\{4\}}} = \widetilde{\beta}_{\{1,4\}}^{\{3,4\}}, \label{eq:apa b14} \\
  \text{for} \ B_{2,\{3\}} &: \quad \alpha_{4,\{3\}}x_{4,2}=\beta^{\{3,4\}}_{\{2,3\}}\alpha_{2,\{3\}}
\\& \Longleftrightarrow  \frac{\beta^{\{3,4\}}_{\{2,3\}}}{x_{4,2}} = \frac{\alpha_{4,\{3\}}}{\alpha_{2,\{3\}}} = \widetilde{\beta}_{\{2,3\}}^{\{3,4\}}, \label{eq:apa b23} \\
  \text{for} \ B_{2,\{4\}} &: \quad \alpha_{3,\{4\}}x_{3,2}=\beta^{\{3,4\}}_{\{2,4\}}\alpha_{2,\{4\}}
\\& \Longleftrightarrow  \frac{\beta^{\{3,4\}}_{\{2,4\}}}{x_{3,2}} = \frac{\alpha_{3,\{4\}}}{\alpha_{2,\{4\}}} = \widetilde{\beta}_{\{2,4\}}^{\{3,4\}}. \label{eq:apa b24}
\end{align}
In Fig.~\ref{fig: k=4 r=2 t=1:spanning} (recall we did not write the superscript $\{3,4\}$), starting with $\widetilde{\beta}_{\{3,4\}}^{\{3,4\}} = \beta_{\{3,4\}}^{\{3,4\}} = -1$, we arrive at the hierarchy~1 $\widetilde{\beta}^{\{3,4\}}_{\{\ell,j\}}, \ell \in \{1,2\}, j \in \{3,4\}$ through $\mathsf{c}^{\{3,4\}}_{\{3\}} = \alpha_{4,\{3\}}$ and $\mathsf{c}^{\{3,4\}}_{\{4\}} = \alpha_{3,\{4\}}$.

Next we have 
\begin{align}
  \text{for} \ {\green B_{1,\{1\}}} &: \quad 0=\beta^{\{3,4\}}_{\{1,3\}}\alpha_{3,\{1\}}x_{3,1}+\beta^{\{3,4\}}_{\{1,4\}}\alpha_{4,\{1\}}x_{4,1} 
\\& \Longleftrightarrow  \frac{\alpha_{4,\{3\}}\alpha_{3,\{1\}}}{\alpha_{1,\{3\}}}+\frac{\alpha_{3,\{4\}}\alpha_{4,\{1\}}}{\alpha_{1,\{4\}}}=0, 
  \label{eq:apa magic1}
  \\
  \text{for} \ {\magenta B_{2,\{2\}}} &: \quad 0=\beta^{\{3,4\}}_{\{2,3\}}\alpha_{3,\{2\}}x_{3,2}+\beta^{\{3,4\}}_{\{2,4\}}\alpha_{4,\{2\}}x_{4,2} 
\\& \Longleftrightarrow  \frac{\alpha_{4,\{3\}}\alpha_{3,\{2\}}}{\alpha_{2,\{3\}}}+\frac{\alpha_{3,\{4\}}\alpha_{4,\{2\}}}{\alpha_{2,\{4\}}}=0,
  \label{eq:apa magic2}
\end{align}
where~\eqref{eq:apa magic1} and~\eqref{eq:apa magic2} are two cycles in Fig.~\ref{fig: k=4 r=2 t=1:spanning} (starting from $\widetilde{\beta}_{\{3,4\}}^{\{3,4\}}$ along the edge with label $-\alpha_{4, \{3\}}$, one is clockwise and the other is counterclockwise) and impose constraints among the involved encoding coefficients. Furthermore, \eqref{eq:apa magic1} and~\eqref{eq:apa magic2} cover another two vertexes $\mathsf{c}^{\{3,4\}}_{\{1\}}$ and $\mathsf{c}^{\{3,4\}}_{\{2\}}$. Indeed in Fig.~\ref{fig: k=4 r=2 t=1:spanning}, by proceeding from $\widetilde{\beta}_{\{2,4\}}$ along the edge with label $\alpha_{4,\{2\}}$, we get 
\begin{align}
  \mathsf{c}^{\{3,4\}}_{\{2\}} 
    = \frac{\alpha_{3,\{4\}}\alpha_{4,\{2\}}}{\alpha_{2,\{4\}}}
\end{align}
and, from $\widetilde{\beta}_{\{2,3\}}$ along the edge with label $-\alpha_{3,\{2\}}$, we get 
\begin{align}
\mathsf{c}^{\{3,4\}}_{\{2\}} 
  = -\frac{\alpha_{4,\{3\}}\alpha_{3,\{2\}}}{\alpha_{2,\{3\}}}.
\end{align}
Similarly for $\mathsf{c}^{\{3,4\}}_{\{1\}}$. By breaking these cycles we obtain \eqref{eq:ex421:c1} and \eqref{eq:ex421:c2}.

Finally, with the condition in~\eqref{eq:apa magic1} and~\eqref{eq:apa magic2}, we get the  hierarchy~2 decoding coefficients
\begin{align}
\text{for} \ {\blue B_{2,\{1\}}} &: \quad -\beta^{\{3,4\}}_{\{1,2\}}\alpha_{2,\{1\}} 
\\&= \beta^{\{3,4\}}_{\{1,3\}}\alpha_{3,\{1\}}x_{3,2}+\beta^{\{3,4\}}_{\{1,4\}}\alpha_{4,\{1\}}x_{4,2}
\\   &= (-x_{4,1}x_{3,2}+x_{3,1}x_{4,2})\frac{\alpha_{3,\{4\}}\alpha_{4,\{1\}}}{\alpha_{1,\{4\}}}
\\ \Longleftrightarrow \widetilde{\beta}_{\{1,2\}}^{\{3,4\}} 
  &= -\ \frac{\beta^{\{3,4\}}_{\{1,2\}}}{(x_{3,1}x_{4,2} -x_{4,1}x_{3,2})} 
\\&= - \ \frac{\mathsf{c}^{\{3,4\}}_{\{1\}}}{\alpha_{2,\{1\}}} = -\ \frac{\alpha_{3,\{4\}}\alpha_{4,\{1\}}}{\alpha_{1,\{4\}} \alpha_{2,\{1\}}}; 
\label{eq:apa b12-1}
\\
\text{for} \ {\cyan B_{1,\{2\}}} &: \quad -\beta^{\{3,4\}}_{\{1,2\}}\alpha_{1,\{2\}} 
\\&= \beta^{\{3,4\}}_{\{2,3\}}\alpha_{3,\{2\}}x_{3,1}+\beta^{\{3,4\}}_{\{2,4\}}\alpha_{4,\{2\}}x_{4,1}
\\   &=-(x_{3,1}x_{4,2}-x_{4,1}x_{3,2})\frac{\alpha_{3,\{4\}}\alpha_{4,\{2\}}}{\alpha_{2,\{4\}}} 
\\  \Longleftrightarrow \widetilde{\beta}_{\{1,2\}}^{\{3,4\}} 
  &= \frac{\beta^{\{3,4\}}_{\{1,2\}}}{(x_{3,1}x_{4,2} -x_{4,1}x_{3,2})} 
\\&= \frac{\mathsf{c}^{\{3,4\}}_{\{2\}}}{\alpha_{1,\{2\}}} = \frac{\alpha_{3,\{4\}}\alpha_{4,\{2\}}}{\alpha_{2,\{4\}} \alpha_{1,\{2\}}}.
\label{eq:apa b12-2}
\end{align}
Indeed in Fig.~\ref{fig: k=4 r=2 t=1:spanning}, we have two paths that lead to $\widetilde{\beta}_{\{1,2\}}^{\{3,4\}}$: 
(i) by proceeding  from $\mathsf{c}_{\{2\}}$ along the edge with label $\alpha_{1,\{2\}}$ we get to $\widetilde{\beta}^{\{3,4\}}_{\{1,2\}}$ as in~\eqref{eq:apa b12-2}, while
(ii) from $\mathsf{c}_{\{1\}}$ along the edge with label $-\alpha_{2,\{1\}}$ we get $\widetilde{\beta}^{\{3,4\}}_{\{1,2\}}$ as in~\eqref{eq:apa b12-1};
but the two must be equal, thus we get the condition in~\eqref{eq:ex421:c3}.

By combining the conditions in~\eqref{eq:apa magic1}, \eqref{eq:apa magic2}, \eqref{eq:apa b12-1}, and~\eqref{eq:apa b12-2}, we have
\begin{align}
  - \frac{\alpha_{4,\{3\}}}{\alpha_{3,\{4\}}}  
  &= \frac{\alpha_{4,\{1\}}}{\alpha_{1,\{4\}}} \ \frac{\alpha_{1,\{3\}}}{\alpha_{3,\{1\}}}
\\&= \frac{\alpha_{4,\{2\}}}{\alpha_{2,\{4\}}} \ \frac{\alpha_{2,\{3\}}}{\alpha_{3,\{2\}}}
\\&= - \frac{\alpha_{2,\{3\}}\alpha_{4,\{1\}}\alpha_{1,\{2\}}}{\alpha_{3,\{2\}}\alpha_{1,\{4\}}\alpha_{2,\{1\}}}, \label{eq: apa relationship}
\end{align}
\hspace{-3.2pt}which is the same as the relationships~\eqref{eq:ex421:wb12} we obtained from the spanning tree in Fig.~\ref{fig: k=4 r=2 t=1:spanning}.

\section{Example: $\Ksf=5, r=2, t=1$}
\label{example: k=5 r=2 t=1}
WLOG, let $\Lc = \{1,2\}$ and thus $\overline{\Lc} = \{3,4,5\}$.
The multicast messages sent by the server are
\begin{align}
W_{\{1,2\}} &= \alpha_{1,\{2\}}B_{1,\{2\}} + \alpha_{2,\{1\}}B_{2,\{1\}}, \quad \text{(all leaders)}, \label{eq:apb w12}\\
W_{\{1,3\}} &= \alpha_{1,\{3\}}B_{1,\{3\}} + \alpha_{3,\{1\}}B_{3,\{1\}}, \quad \text{(mixed)},\\
W_{\{1,4\}} &= \alpha_{1,\{4\}}B_{1,\{4\}} + \alpha_{4,\{1\}}B_{4,\{1\}}, \quad \text{(mixed)},\\
W_{\{1,5\}} &= \alpha_{1,\{5\}}B_{1,\{5\}} + \alpha_{5,\{1\}}B_{5,\{1\}}, \quad \text{(mixed)},\\
W_{\{2,3\}} &= \alpha_{2,\{3\}}B_{2,\{3\}} + \alpha_{3,\{2\}}B_{3,\{2\}}, \quad \text{(mixed)},\\
W_{\{2,4\}} &= \alpha_{2,\{4\}}B_{2,\{4\}} + \alpha_{4,\{2\}}B_{4,\{2\}}, \quad \text{(mixed)},\\
W_{\{2,5\}} &= \alpha_{2,\{5\}}B_{2,\{5\}} + \alpha_{5,\{2\}}B_{5,\{2\}}, \quad \text{(mixed)},\label{eq:apb w25}
\end{align}
and those that we not sent are
\begin{align}
W_{\{3,4\}} &= \alpha_{3,\{4\}}B_{3,\{4\}} + \alpha_{4,\{3\}}B_{4,\{3\}}, \quad \text{(all non leaders)},\\
W_{\{3,5\}} &= \alpha_{3,\{5\}}B_{3,\{5\}} + \alpha_{5,\{3\}}B_{5,\{3\}}, \quad \text{(all non leaders)},\\
W_{\{4,5\}} &= \alpha_{4,\{5\}}B_{4,\{5\}} + \alpha_{5,\{4\}}B_{5,\{4\}}, \quad \text{(all non leaders)}.
\end{align}

For every $\Ac \in \Omega_{\{3,4,5\}}^{2}$, the non-send multicast message $W_\Ac$ can be reconstruct from $\{W_\Sc : \Sc \in \Omega_{\{1,2\} \cup \Ac}^{2}, \Sc \neq \Ac \}$, by a procedure equivalent to~\eqref{eq:apb w12}-\eqref{eq:apb w25} after appropriate relabeling of the indices of the non-leader users. To locally reconstruct all the non-sent multicast messages we thus proceed as for ``reduced systems'' with parameters $\Ksf^\prime=4, r=2, t=1$. 
By symmetry and from~\eqref{eq: apa relationship}, the relationships among the encoding coefficients are
\begin{align}
  \Ac=\{3,4\} :
  - \frac{{\cyan \alpha_{4,\{3\}}}} {\alpha_{3,\{4\}}}  
  &= \frac{\alpha_{4,\{1\}}}{\alpha_{1,\{4\}}} \ \frac{\alpha_{1,\{3\}}}{\alpha_{3,\{1\}}}
\\&= \frac{\alpha_{4,\{2\}}}{{\cyan \alpha_{2,\{4\}}}} \ \frac{{\cyan \alpha_{2,\{3\}}}}{\alpha_{3,\{2\}}}
\\&= - \frac{{\cyan \alpha_{2,\{3\}}}\alpha_{4,\{1\}}\alpha_{1,\{2\}}}{\alpha_{3,\{2\}}\alpha_{1,\{4\}}\alpha_{2,\{1\}}},
    \label{eq: apb w34}
\\
  \Ac=\{3,5\} :
  - \frac{{\cyan \alpha_{5,\{3\}}}}{\alpha_{3,\{5\}}}  
  &= \frac{\alpha_{5,\{1\}}}{\alpha_{1,\{5\}}} \ \frac{\alpha_{1,\{3\}}}{\alpha_{3,\{1\}}}
\\&= \frac{\alpha_{5,\{2\}}}{{\cyan \alpha_{2,\{5\}}}} \ \frac{{\cyan \alpha_{2,\{3\}}}}{\alpha_{3,\{2\}}}
\\&= - \frac{{\cyan \alpha_{2,\{3\}}}\alpha_{5,\{1\}}\alpha_{1,\{2\}}}{\alpha_{3,\{2\}}\alpha_{1,\{5\}}\alpha_{2,\{1\}}},
  \\
  \Ac=\{4,5\} :
  - \frac{{ \cyan \alpha_{5,\{4\}}}}{\alpha_{4,\{5\}}}  
  &= \frac{\alpha_{5,\{1\}}}{\alpha_{1,\{5\}}} \ \frac{\alpha_{1,\{4\}}}{\alpha_{4,\{1\}}}
\\&= \frac{\alpha_{5,\{2\}}}{{\cyan \alpha_{2,\{5\}}}} \ \frac{{\cyan \alpha_{2,\{4\}}}}{\alpha_{4,\{2\}}}
\\&= - \frac{{\cyan \alpha_{2,\{4\}}}\alpha_{5,\{1\}}\alpha_{1,\{2\}}}{\alpha_{4,\{2\}}\alpha_{1,\{5\}}\alpha_{2,\{1\}}},
\end{align}
where the coefficients highlighted in cyan are assigned to dotted edges in Fig.~\ref{fig: k=5 r=2 t=1}.
For example, for $\Ac=\{3,4\}$ (and similarly for $\Ac=\{3,5\}$ and $\Ac=\{4,5\}$), the relationships revealed in Fig.~\ref{fig:521-w34} are
\begin{align}
  {\cyan \alpha_{4,\{3\}}} = -\ \frac{\alpha_{4, \{1\}} \alpha_{3, \{4\}} \alpha_{1, \{3\}}}{\alpha_{3, \{1\}} \alpha_{1, \{4\}}}, \label{eq: a43}\\
  {\cyan \alpha_{2,\{4\}}} = -\ \frac{\alpha_{2, \{1\}} \alpha_{4, \{2\}} \alpha_{1, \{4\}}}{\alpha_{4, \{1\}} \alpha_{1, \{2\}}}, \label{eq: a24}\\
  {\cyan \alpha_{2,\{3\}}} = -\ \frac{\alpha_{2, \{1\}} \alpha_{3, \{2\}} \alpha_{1, \{3\}}}{\alpha_{3, \{1\}} \alpha_{1, \{2\}}}. \label{eq: a23}
\end{align}
By substituting the fixed coefficients in \eqref{eq: a43}-\eqref{eq: a23} into~\eqref{eq: apb w34}, we will eliminate other free coefficients, that is, \eqref{eq: apb w34} are equivalent to~\eqref{eq: a43}-\eqref{eq: a23}, which we obtained from the spanning trees in Fig.~\ref{fig: k=5 r=2 t=1} by using the greedy algorithm in Section~\ref{sec:many leader sets}.

\bibliographystyle{IEEEtran}
\bibliography{references}

\begin{thebibliography}{1}
\providecommand{\url}[1]{#1}
\csname url@samestyle\endcsname
\providecommand{\newblock}{\relax}
\providecommand{\bibinfo}[2]{#2}
\providecommand{\BIBentrySTDinterwordspacing}{\spaceskip=0pt\relax}
\providecommand{\BIBentryALTinterwordstretchfactor}{4}
\providecommand{\BIBentryALTinterwordspacing}{\spaceskip=\fontdimen2\font plus
\BIBentryALTinterwordstretchfactor\fontdimen3\font minus
  \fontdimen4\font\relax}
\providecommand{\BIBforeignlanguage}[2]{{%
\expandafter\ifx\csname l@#1\endcsname\relax
\typeout{** WARNING: IEEEtran.bst: No hyphenation pattern has been}%
\typeout{** loaded for the language `#1'. Using the pattern for}%
\typeout{** the default language instead.}%
\else
\language=\csname l@#1\endcsname
\fi
#2}}
\providecommand{\BIBdecl}{\relax}
\BIBdecl

\bibitem{maddah2014fundamental}
M.~A. Maddah-Ali and U.~Niesen, ``Fundamental limits of caching,'' \emph{IEEE
  Transactions on Information Theory}, vol.~60, no.~5, pp. 2856--2867, 2014.

\bibitem{yu2017exact}
Q.~Yu, M.~A. Maddah-Ali, and A.~S. Avestimehr, ``The exact rate-memory tradeoff
  for caching with uncoded prefetching,'' \emph{IEEE Transactions on
  Information Theory}, vol.~64, no.~2, pp. 1281--1296, 2017.

\bibitem{wan2020index}
K.~Wan, D.~Tuninetti, and P.~Piantanida, ``An index coding approach to caching
  with uncoded cache placement,'' \emph{IEEE Transactions on Information
  Theory}, vol.~66, no.~3, pp. 1318--1332, 2020.

\bibitem{wan2020cache}
K.~Wan, H.~Sun, M.~Ji, D.~Tuninetti, and G.~Caire, ``Cache-aided scalar linear
  function retrieval,'' in \emph{2020 IEEE International Symposium on
  Information Theory (ISIT)}.\hskip 1em plus 0.5em minus 0.4em\relax IEEE,
  2020, pp. 1717--1722.

\bibitem{sun2018capacity}
H.~Sun and S.~A. Jafar, ``The capacity of private computation,'' \emph{IEEE
  Transactions on Information Theory}, vol.~65, no.~6, pp. 3880--3897, 2018.

\bibitem{yu2018characterizing}
Q.~Yu, M.~A. Maddah-Ali, and A.~S. Avestimehr, ``Characterizing the rate-memory
  tradeoff in cache networks within a factor of 2,'' \emph{IEEE Transactions on
  Information Theory}, vol.~65, no.~1, pp. 647--663, 2018.

\end{thebibliography}

\end{document}